# Impact of genetic polymorphisms on tacrolimus concentrations and intra-individual variability in recipients of heart transplants during the early post-heart transplantation period

**Authors:** Yuhui Chai[1#], Lili Hu[2#], Danni Quan[1], Yunyun Yang[1*], ZhuoWang[1*]

**ORCIDs:** Yuhui Chai: https://orcid.org/0009-0008-3751-8643, Lili Hu: https://orcid.org/0000-0001-8861-5672, Danni Quan: https://orcid.org/0009-0003-9779-4635, Yunyun Yang: https://orcid.org/0000-0002-0395-3662, ZhuoWang: https://orcid.org/0000-0003-1252-7961.

**Affiliations:**

[1]Department of Pharmacy, Shanghai Changhai Hospital, the First Affiliated Hospital of Navy Medical University, Shanghai, 200433, China

[2]Department of Pharmacy, 904th Hospital, Joint Logistics Support Force, Wuxi 214000, Jiangsu Province, China

**\*Corresponding author:** Yunyun Yang[1], ZhuoWang[1]; email: 13262246965@163.com (Yunyun Yang), wztgyx223@163.com (Zhuo Wang).

[#]These authors equally contributed to this study

**Abbreviations:**

ANOVA, analysis of variance; $C_0$, trough concentration; CNI, calcineurin inhibitors; CV, coefficient of variation; CYP; cytochrome P450; IL, interleukin; IPV, intra-individual variability; PD, pharmacodynamics; PK, pharmacokinetics; SD, standard deviation; SNP, single nucleotide polymorphism; Tac, tacrolimus; TDM, therapeutic drug monitoring;

**Word count:** Main text and abstract: 3,986 words

**Abstract**

This study aimed to investigate the effects of genetic polymorphisms on tacrolimus blood levels and intra-individual variability in recipients of heart transplants during the early post-transplantation period. Demographic information, concomitant medications, daily tacrolimus dose, trough concentration, and physiological and biochemical information of 87 Chinese recipients of heart transplants were collected. Trough concentrations were determined using a chemiluminescent micro-particle immunoassay, and 17 selected single nucleic acid polymorphisms were genotyped by direct sequencing. We assessed intra-individual variability by calculating the coefficient of variation of tacrolimus trough concentration and analyzed factors associated with tacrolimus concentration and intra-individual variability. Our study found that low body weight and a high percentage of neutrophils significantly influenced the coefficient of variation of tacrolimus. CYP3A5*1D and CYP3A7 rs776744 haplotypes correlated significantly with an intra-individual coefficient of variation of tacrolimus trough concentration during the early postoperative period. Patients with the CYP3A5*1D rs15524 and CYP3A7 rs776744 TT haplotype had a higher coefficient of variation than carriers of the C allele. Genetic polymorphisms in recipients of heart transplants affect tacrolimus metabolism, significantly affecting tacrolimus blood concentration during the early postoperative period. Genotyping before drug administration can help optimize dosing regimens, reduce intra-individual variability, and improve prognosis.

# 1 INTRODUCTION

Heart transplantation is the ultimate treatment for most patients with end-stage heart failure. Since the first heart transplantation in 1967, survival rates have improved considerably, with an increase in the median survival for adult heart transplantations now at 11 years.[1] Calcineurin inhibitors (CNI) are central to immunosuppressive therapy following heart transplantation. Tacrolimus (Tac), a macrolide CNI with potent immunosuppressive properties and a lower incidence of side effects than cyclosporine, has been widely used to prevent post-transplant rejection.[2] Tac mainly forms a complex with Tac-binding protein 12 (FKBP12) in T lymphocytes, which binds to calmodulin phosphatase to inhibit its activity, affecting the expression of cytokines, such as interleukin-2 (IL-2), thus inhibiting T lymphocyte proliferation and differentiation.[3]

Although Tac is crucial for maintaining immunosuppression during heart transplantation, its narrow therapeutic window and large intra- and inter-individual pharmacokinetic (PK) differences predispose patients to rejection from insufficient drug exposure or toxicity caused by excessive drug accumulation.[4] Therefore, early therapeutic drug monitoring (TDM) is essential to optimize the administered dose of Tac and improve postoperative outcomes. Currently, the initial dose of tacrolimus is based on the patient's weight and subsequent adjustments made through TDM. However, some patients struggle to maintain trough concentration ($C_0$) within the target concentration range even after dose modification. Using only the instantaneous concentration of Tac at a given time point to assess drug exposure and treatment response has limited utility.

Currently, there is no consensus on the optimal target concentration of Tac. Recently, the intra-individual variability of Tac (Tac-IPV) has been recognized as a novel prognostic marker for organ transplantation for identifying adverse prognostic risks, such as rejection and graft loss.[5]

Tac-IPV may be better for assessing the risk of adverse outcomes owing to over- or under-exposure to Tac.[6] Coefficient of variation (CV) was also used to assess IPV, calculated by dividing the standard deviation (SD) of the number of serial pre-dose concentrations by the mean Tac measurements.[7] High Tac-IPV is a notable prognostic factor for graft function attributable to T cell- and humoral-mediated rejection as well as vascular changes.[8] In addition, Tac-IPV significantly correlated with adverse prognosis (such as rejection, infection, and death) after heart transplantation.[9,10] Many factors can affect the PK parameters and individual variability of Tac, including methods of Tac concentration analysis, drug interactions, food effects on drug metabolism, the interval between administration and blood collection time, and genetic polymorphisms of drug metabolism in individuals.[11,12] Previous studies focused on patients with stable IPV for at least 3 months after transplantation. However, during the early post-transplant period, the physiological state of the patient's intestine, liver, and whole body is still unstable, causing substantial fluctuations in Tac concentration, and the factors affecting Tac-IPV remain unclear.

Currently, there is an increasing interest in identifying and validating genetic variations that contribute to IPV.[13] Polymorphisms in genes encoding drug-metabolizing enzymes and drug transporters are among the most important factors contributing to the altered activity of Tac, leading to Tac-IPV generation. Tacrolimus is mainly metabolized by the cytochrome P450 (CYP)-3A enzyme system, of which CYP3A4 and CYP3A5 are the most important isoenzymes involved.[14] Additionally, the effects of polymorphisms in other drug-metabolizing enzymes (CYP3A7 rs2257401, CYP3A7 rs10211, and CYP3A7 rs776744), signaling factors (STAT3 rs1053004, SUMO4 rs237024, POR rs2868177, MAP3K9 rs8006424, ADRB2 rs1042713, and

MAP3K1 rs62355944), regulatory factors (PPARA rs5767743), and cytokines (TLR9 rs352139, IL17A rs2275913, and IL3 rs181781) on Tac metabolism in recipients of cardiac transplant remains unclear. Recently, numerous studies have shown that TDM combined with genetic polymorphisms of PK/pharmacodynamic (PD)-related targets can more accurately predict the individual variability of the initial Tac dose and concentration, thereby improving heart transplant prognosis.

Additionally, few studies have revealed the PK of Tac during the early stages of heart transplantation. Domestic researchers have shown that maintaining the target therapeutic window of Tac at 15 – 20 ng/ml within 1 month after heart transplantation is safer and more effective.[15] International guidelines and experts agree that the target concentration of Tac should be 10–15 ng/ml within 3 months post-transplant.[16,17] Owing to the lack of higher-quality evidence-based medicine, the recommended range in the current guidelines for Tac $C_0$ in the early stage after heart transplantation remained unchanged for nearly 20 years.

Accordingly, this study aimed to retrospectively analyze the rate of Tac $C_0$ in the early postoperative period after heart transplantation and investigate its related influencing factors to provide a reference for individualized drug therapy.

## 2 Materials and methods

### 2.1 Study design and population

Our study included 87 patients who underwent heart transplantation at our center between March 2017 and February 2023. All patients received triple-drug immunosuppression (Tac, mycophenolate mofetil, and corticosteroids), with regular monitoring of Tac $C_0$. The inclusion criteria were as follows: Chinese recipients of heart transplants, first-time recipients of heart transplants, and no other organ transplant experience. The exclusion criteria were as follows: incomplete dosing regimen and timing of blood collection, patients who experienced acute rejection, patients who are pregnant or lactating, patients undergoing hemodialysis or peritoneal dialysis during the study period, and patients who had changed or co-administered other immunosuppressive agents. This study was approved by the institutional review board.

2.2 Data collection

Demographic information and laboratory data of the recipients of heart transplants were collected retrospectively from electronic medical records. Data recorded included the recipient's weight (kg), daily dose (mg/day), white blood cell (WBC) count, and other laboratory data on days 7, 14, 21, and 28.

2.3 Tac trough levels and CV

For each patient, whole-blood Tac steady-state (at least 48 h) and trough levels at 1 month post-heart transplantation were documented. Whole blood $C_0$ of Tac was assayed using an Architect Tac Reagent Kit (Abbott, Abbott Park, IL, USA). The assay was a chemiluminescent micro-particle immunoassay routinely used for the quantitative determination of Tac-$C_0$. The CV was calculated as CV = SD/Mean*100%. (SD is the standard deviation of all tested concentrations within 1 month after transplantation, and the mean is the mean of all tested concentrations). Based on the variability in the entire cohort, patients were categorized into high (above the median CV) and low Tac variability (below the median CV) groups.

2.4 Genotyping

Based on the gene loci and potential gene loci affecting the reported Tac PK/PD, we screened 17 single nucleic acid gene loci associated with Tac PK/PD: drug-metabolizing enzymes

(CYP3A5*3, CYP3A4*1G, CYP3A5*1D, CYP3A7 rs2257401, CYP3A7 rs10211, and CYP3A7 rs776744), signaling factors (STAT3 rs1053004, SUMO4 rs237024, POR rs2868177, MAP3K9 rs8006424, ADRB2 rs1042713, and MAP3K1 rs62355944), regulatory factors (POR rs2868177, PPARA rs5767743), and cytokines (TLR9 rs352139, IL17A rs2275913, and IL3 rs181781). Genomic DNA was isolated from the peripheral leukocytes of blood samples using kits. Selected single nucleotide polymorphisms (SNPs) were genotyped by direct sequencing. Genotyping was performed using a reagent vendor.

2.5 Statistical Analysis

The distribution of continuous data was evaluated using the Kolmogorov–Smirnov test; subsequently, parametric or nonparametric tests were applied as appropriate. Results conforming to normal distribution were expressed as Mean ± SD, and differences between groups were compared using a one-sample t-test or one-way analysis of variance (ANOVA). Numerical variables with non-normally distributed results were statistically described as the median and interquartile range (IQR), and the Mann-Whitney U test was used to compare data among groups. Genotype frequencies of the polymorphisms were tested for deviations from Hardy-Weinberg equilibrium using appropriate chi-square tests. One-way ANOVA and multifactorial logistic regression analyses were performed to determine the factors influencing the differences in Tac blood concentrations and IPV. A $P$-value < 0.05 was considered statistically significant. SPSS (version 21.0) and GraphPad Prism (version 8.0) were used for the statistical analyses.

3 Results

3.1 Patient characteristics

The demographic and clinical characteristics of the 87 recipients of heart transplants are shown in Table 1. Among these, 67 (77 %) were male. The average age of the recipients was 47 (range 12 – 73) years, and the average weight was almost 61 kg (61.40 ± 11.88). Based on the guideline for the target concentration within 3 months post-transplantation (10 – 15 ng/mL), 87 patients were categorized into three groups: $C_0 < 10.0$ ng·mL$^{-1}$ (n = 33), $10 \leq C_0 \leq 15$ ng·mL$^{-1}$ (n = 41), and $C_0 > 15$ ng·mL$^{-1}$ (n = 13). As shown in Table 2, sex, age, weight, WBC, neutrophils, hematocrit, hemoglobin, globulin, alanine aminotransferase, and co-medication were not significantly correlated with the Tac $C_0$ distribution.

3.2 Genotype frequencies and Hard-Weinberg equilibrium test

The distributions of CYP3A5*3, CYP3A4*1G, CYP3A5*1D, CYP3A7 rs2257401, CYP3A5 rs4646453, CYP3A7 rs10211,CYP3A7 rs776744,STAT3 rs1053004, SUMO4 rs237024, POR rs2868177, MAP3K9 rs8006424,ADRB2 rs1042713, MAP3K1 rs62355944, PPARA rs5767743, TLR9 rs352139, IL17A rs2275913, and IL3 rs181781 are shown in (Supplementary Table 1). These results are consistent with the Hardy–Weinberg equilibrium.

3.3 Distribution of Tac $C_0$ after heart transplantation

In this study, the Tac $C_0$ was detected 346 times in 87 patients after heart transplantation, yielding a compliance rate of 37.3%. The average daily dose of Tac was (1.99 ± 0.77) mg/kg, and the average $C_0$ of Tac was (7.76 ± 0.83) ng/ml, with a fluctuation range of 0.3 – 17.2 ng/ml, with significant correlation. Tac $C_0$ distribution levels after transplantation are shown in Table 3. Tac $C_0$/D is the ratio of trough concentration to daily dose adjusted for body weight, commonly used to evaluate Tac metabolism. The average Tac $C_0$/D value at different time points after operation: D7: 4.57 ng·mL-1, D14: 8.34 ng·mL-1, D21: 9.34 ng·mL-1, D28: 8.24 ng·mL-1. The $C_0$/D ratio on D7 was the lowest and significantly different from that of the other periods, as shown in Figure 1.

3.4 Univariate analysis of factors influencing the CV of Tac

To further explore the factors influencing Tac-IPV after heart transplantation, we used the classical CV calculations. The 87 recipients were categorized into the high (CV ≥ 0.38) and low IPV group (CV < 0.38) based on the median CV (0.38). The results were analyzed using independent sample t-tests, and significant differences were observed between the two groups in terms of body weight, daily dose of Tac, percentage of neutrophils, and hemoglobin concentration (P<0.05) (Table 4).

3.4.1 Multivariate logistic regression analysis of CV of Tac

Statistically significant factors in univariate analysis were included in the multivariate logistic regression model, and the CV < 0.38 group was used as the control. Logistic analysis showed that low body weight (OR (odds ratio) = 0.95，95%：0.920 - 0.998) and high percentage of neutrophils (OR = 4.404，95%：1.53 - 12.616) significantly affect the CV of Tac (Table 5). This indicated that neutrophil weight and percentage may be independent risk factors.

3.5 Effect of gene polymorphism on Tac metabolism in the early stage post-heart transplantation

3.5.1 Correlation between gene polymorphisms and Tac $C_0$ distribution in recipients of heart transplants

We examined 17 SNPs to explore the correlation between gene polymorphisms and Tac $C_0$ distribution in recipients of heart transplants. We found that only the CYP3A4*1G genotype was significantly correlated with Tac $C_0$ distribution (Supplementary Table 2). The proportion of CYP3A4 *1G genotype carriers decreased progressively with increasing Tac $C_0$ (Figure 2).

3.5.2 Influence of gene polymorphisms on Tac $C_0$/D in the early postoperative period

To minimize the influence of the administered dose, body weight, and postoperative recovery process on Tac $C_0$, this study analyzed Tac $C_0$ in week 4 post-transplantation, with adjustments for dose and body weight. The influences of the 17 SNPs on the Tac $C_0/D$ ratio are shown in Supplementary Table 3. CYP3A5*3 (Figure 3A), CYP3A5 rs4646453 (Figure 3B), CYP3A*1D (Figure 3C), CYP3A7 rs10211 (Figure 3D), CYP3A7 rs776744 (Figure 3E), CYP3A7 rs2257401 (Figure 3F), and IL3 rs181781 (Figure 3G) were significantly associated with Tac $C_0/D$ during the early post-transplantation period ($P <0.05$). Among the drug-metabolizing enzyme-related genes, CYP3A5*3 GG type had higher $C_0/D$ values than carriers of A allele, CYP3A5 rs4646453 GG type had higher $C_0/D$ values than A allele carriers, CYP3A5*1D TT type had higher $C_0/D$ values than C allele carriers, CYP3A7 rs2257401 CC type had higher $C_0/D$ values than G allele carriers, CYP3A7 rs10211 AA type had higher $C_0/D$ values than G allele carriers, and CYP3A7 rs776744 TT type have higher $C_0/D$ values than C allele carriers. Among the cytokine-related genes, the IL3 rs181781 AA genotype had higher $C_0/D$ values than G allele carriers.

3.5.3 Effect of gene polymorphisms on CV of Tac in the early postoperative period after heart transplantation

In our study, Tac-IPV was assessed by calculating its CV. We explored the effect of different genotypes on the CV of Tac and showed that the CYP3A5*1D and CYP3A7 rs776744 SNPs were significantly correlated with the CV of Tac in the early postoperative period after heart transplantation ($P <0.05$). Patients with CYP3A5*1D and CYP3A7 rs776744 TT-type had higher CV values than carriers of the C allele (Table 6).

## 4 Discussion

As a first-line anti-rejection drug in heart transplantation, Tac has a narrow therapeutic window, which is greatly affected by drug combination and gene polymorphisms, resulting in considerable variability among individuals. The 30-day postoperative period is critical for the early recovery of recipients of heart transplants, during which they experience high intra-and inter-individual variability in Tac owing to the presence of infections and other high-risk factors, posing a great challenge in rationalizing the use of Tac. In our study, we retrospectively analyzed post-transplantation Tac $C_0$ in 87 recipients of cardiac transplants to identify the factors affecting these concentrations and elucidated factors producing high IPV to provide a relevant basis for dosage adjustment of Tac in the early stage of post-transplantation and to identify the population with high IPV. Our study findings showed that the Tac compliance rate in the early postoperative period after heart transplantation was only 37.3 %, and the mean Tac $C_0$ was below the target. A significant difference was observed in $C_0/D$ among the patients at different times postoperatively, suggesting that there was a difference in the ability to metabolize FK506 at different times. In addition, we found that postoperative co-administration of voriconazole in recipients of heart transplants caused an increase in Tac $C_0$. Voriconazole may block the elimination of Tac by inhibiting the CYP3A4 enzyme and P-gp activity, resulting in drug accumulation and increased Tac levels, consistent with previous reports.[18] Although Tac $C_0$ monitoring is a routine assessment after various organ transplants, inadequate management of its concentration often leads to poor correlation between Tac $C_0$ and clinical prognosis indicators. Recently, the association between Tac-IPV and long-term graft outcomes has become widely accepted in transplantation. CV is a widely used method for calculating the Tac-IPV. This ratio, obtained by dividing the SD by the mean, is used as a measure of the dispersion of the data.[19] The stability of the Tac $C_0$ was assessed using CV.[20,21] In the high and low CV groups, weight, daily dose of Tac, percentage of neutrophils, and hemoglobin concentration were the main factors affecting Tac-IPV, and logistic regression analysis showed that lower body weight and higher percentage of neutrophils were independent risk factors for high Tac-IPV. This finding is the first to suggest a new indicator affecting Tac-IPV, following drug interactions, patient compliance, and drug use time. Musuamba et al.[22] showed that weight affects not only metabolism but also the apparent distribution of Tac. Considering that Tac administration is based on the patient's weight, fluctuations in weight can lead to variations in drug concentration. Tac is widely bound to red blood cells (RBCs), and changes in hemoglobin concentration affect the distribution of tacrolimus in RBCs and plasma.[23] Although changes in neutrophils may be associated with infection or inflammation, these conditions can affect the metabolic abilities of patients, indirectly affecting the change in Tac CV values.

TDM results can exhibit delay, making it challenging to achieve an initial dosing regimen in some patients who are fast-metabolizers, resulting in low and fluctuating $C_0$ values in the early patients. This suggests that the initial dose should be adjusted based on the patient's metabolic ability. CYP3A5 and CYP3A4 are the main enzymes involved in Tac metabolism. Among these, CYP3A5*3 has a high mutation frequency in the population, which significantly correlated with Tac elimination. Consequently, this genotype has been recognized as a biomarker of the Tac metabolic phenotype in most studies.[24] Studies have shown that CYP3A5 gene polymorphism is associated with the PK of Tac, and other studies have suggested that the CYP3A5*3 gene polymorphism may not correlate with Tac $C_0$ within 1 month after heart transplantation.[25] Our results showed that CYP3A5*3, CYP3A5*1D, and CYP3A5 rs4646453 genotypes significantly correlated with $C_0$/D values in heart transplant recipients. Specifically, the CYP3A5*3 GG type had higher $C_0$/D values than the A allele carriers, further validating that the CYP3A5*3 allele is associated with Tac PK. CYP3A5 1D is involved in Tac metabolism after renal transplantation and in stable renal transplant recipients, where it significantly influenced $C_0$/D values of Ta and sirolimus at different times from 4 to 24 months after transplantation.[26] CYP3A5 1D is involved in Tac metabolism after liver transplantation. On the 3rd day, 1st and 2nd week, 1st, 2nd, and 3rd month after liver transplantation, the Tac $C_0$/D of TT genotype carriers of CYP3A5 1D was higher than that of patients with the CT and CC alleles.[27] Our results are consistent with previous studies; CYP3A5 1D TT genotype carriers had higher Tac $C_0$/D values. In addition, we observed that CYP3A5 rs4646453 was involved in Tac metabolism after heart transplantation, a result consistent with findings in kidney transplantation [28].

However, the influence of CYP3A4 polymorphisms on Tac-IPV remains controversial. CYP3A4*1G, one of the SNP with the highest mutation rates in the CYP3A4 SNP family, has been studied in a Chinese population.[29] Our study showed that the Tac $C_0$ in CYP3A4 *1G carriers was lower than the effective concentration, leading to slightly poorer clinical outcomes. This may be attributed to mutations at this CYP3A 4 *1G gene, which increase the activity of the CYP3A4 enzyme, thereby accelerating the metabolism of TAC.

CYP3A7 was initially thought to be expressed only during fetal development; however, recent studies have detected high expression of CYP3A7 in more than half of adult livers, contributing approximately 20% of the total CYP3A.[30] Sun et al. found that CYP3A7 rs2257401 and rs10211 were involved in cyclosporine metabolism at different times after kidney transplantation in China.[31] A whole-exome sequencing study showed that CYP3A7 rs2257401 was associated with Tac $C_0/D$ in adult recipients of kidney transplants.[32] These findings suggest that CYP3A7 polymorphisms may serve as important genetic markers of Tac PK in transplant patients. Similarly, our study found that CYP3A7 rs776744, CYP3A7 rs2257401, and CYP3A7 rs10211 polymorphisms were significantly correlated with TAC $C_0/D$, with CYP3A7 rs776744 T, CYP3A7 rs2257401 C, and CYP3A7 rs10211 A gene carriers having significantly higher $C_0/D$ than non-carriers. Small ubiquitin-associated modifier protein 4 (SUMO4) is located in the cytoplasm and negatively regulates NF-κB, which in turn affects the expression of retinoid X receptor-regulated CYP3A4 and CYP3A5.[33] This ultimately leads to individual differences in Tac PK of tacrolimus.[34] However, our study did not find a significant association between SUMO4 rs237024 and Tac metabolism in the early postoperative period after heart transplantation, as there was no significant difference in $C_0/D$ values between SUMO4 rs237024 CC-type carriers and non-carriers. Owing to differences in transplantation type and races, regulation of NF-κB by SUMO4 may affect Tac metabolism through other mechanisms or pathways.

In addition, IL-3 and its SNPs, IL-3 rs181781, play roles in immune regulation and cell signal transduction, particularly in acute rejection after renal transplantation.[35] Previous studies reported that patients with the IL-3 rs181781 AA genotype have significantly higher Log $C_0/D$ values than patients with the AG and GG genotypes at 30 and 90 days after kidney transplantation.[36] Similarly, our results showed that the $C_0/D$ value of the IL-3 rs181781 AA genotype was significantly higher than that of the AG and GG genotype patients 30 days after heart transplantation, and the required dosage was also lower. This study is the first to investigate the relationship between the IL-3 rs181781 polymorphism and Tac metabolism in Chinese heart transplant recipients.

Recently, Tac-IPV emerged as a novel marker to assess the risk of adverse outcomes in solid organ transplantation.[37] Tac-IPV measures the fluctuation of $C_0$ within a patient over a specific period, which is quantified using various metrics, such as SD and CV. In this study, we used the classic CV calculation method to calculate Tac-IPV and investigate the influence of gene polymorphisms. Our study demonstrated the statistically significant effects of CYP3A5*1D and CYP3A7 rs776744 SNPs on Tac-IPV. Patients with CYP3A5*1D and CYP3A7 rs776744 TT phenotypes showed higher Tac-IPV than those with other genotypes. More research on pharmacogenomics and Tac-IPV has been carried out on the CYP3A5*3 genotype; however, the influence of CYP3A5*3 gene polymorphisms on Tac-IPV remains controversial. A study in South Korea showed that Tac-IPV of CYP3A5 *3/*3 genotype was higher than that of CYP3A5 *1/*1 and *1/*3 groups by >50%.[38] However, similar associations between the CYP3A5 genotypes and Tac-IPV were not reported in other studies.[39,40] In this study, we did not observe any correlation between CYP3A5 and Tac-IPV. The influence of CYP3A5 gene polymorphisms on Tac-IPV needs to be further clarified because the calculation methods of IPV, transplant time, and condition of the transplanted kidney are different. Our results confirmed that the SNPs of CYP3A5*1D and CYP3A7 rs776744 were significantly correlated with Tac-IPV after heart transplantation, suggesting that Tac-IPV can be reduced from the perspective of gene polymorphisms to improve transplantation outcomes.

Because tacrolimus concentrations do not reach steady-state levels within 1 week after heart transplantation, and the patient's pathophysiology is unstable within 2 weeks, these factors may obscure the impact of genetic factors. In contrast, with extended transplantation time, changes in the glucocorticoid dose can cause changes in the CYP enzyme, affecting the judgment of the results. Therefore, we finally included Tac $C_0$ at 1 month (30 ± 3 d) after transplantation. These results are limited to the early stages after heart transplantation and need to be further verified in large-sample multicenter studies. The potential limitations of this study should be considered when interpreting the results. The effect of post-transplant gastrointestinal dysfunction on Tac $C_0$-level and Tac-IPV cannot be ruled out.

In conclusion, our study found that the mean Tac $C_0$ was lower than the target concentration range in the early stages after heart transplantation; however, its correlation with prognosis (such as rejection) remains under-explored. Further research is needed to explore whether the target Tac concentration range needs to be reduced in the early post-transplantation period. This is the first study to report that weight, daily Tac dose, percentage of neutrophils, and hemoglobin concentration are the main factors affecting Tac-IPV. Moreover, low body weight and higher neutrophil percentage are independent risk factors for high Tac-IPV, whereas the SNPs of CYP3A5*1D and CYP3A7 rs776744 were significantly related to Tac-IPV. These results showed that genotyping and drug concentration can be used objectively to predict dose for individuals with different genotypes, thus improving the efficacy of immunotherapy, reducing IPV, and improving prognosis.

**Acknowledgements/ Funding:** We would like to thank Editage (www.editage.cn) for English language editing.

**Disclosure:** All other authors of this manuscript have no conflicts of interest to disclose asdescribed by the American Journal of Transplantation.

**Data availability statement:** The data are not publicly available due to (restrictions e.g., their containing information that could compromise the privacy of research participants).

**Declaration of Generative AI and AI-assisted technologies in the writing process:** None

**Figure legends:**

Figure 1. The change in mean $C_0/D$ of Tac after heart transplantation. Data were presented as mean ± SEM. *$p < 0.05$ vs. the D7 group. Tac: tacrolimus; $C_0$: blood trough TAC concentration; $C_0/D$: dose-adjusted trough blood concentrations.

Figure 2. Characterization of genotype distribution of CYP3A4*1G single nucleic acid polymorphisms at different Tac blood concentration distributions. Tac: tacrolimus.

Figure 3. Influence of different genotypes on the Tac $C_0/D$ ratio. (**A**) Effect of CYP3A5 gene polymorphisms on Tac $C_0/D$ in the early stage after heart transplantation. *$p < 0.05$ vs. the CYP3A5*3 GG group. (**B**) Effect of CYP3A5 rs4646453 gene polymorphisms on Tac $C_0/D$ in the early stage after heart transplantation. *$p < 0.05$ vs. the CYP3A5 rs4646453 GG group. (**C**) Effect of CYP3A5*1D gene polymorphisms on Tac $C_0/D$ in the early stage after heart transplantation. *$p < 0.05$ vs. the CYP3A5 *1D CC group. (**D**) Effect of CYP3A7 rs10211 gene polymorphisms on Tac $C_0/D$ in the early stage after heart transplantation. *$p < 0.05$ vs. the CYP3A7 rs10211 AA group. (**E**) Effect of CYP3A7 rs776744 gene polymorphisms on Tac $C_0/D$ in the early stage after heart transplantation. *$p < 0.05$ vs. the CYP3A7 rs776744 CC group. (**F**) Effect of CYP3A7 rs2257401 gene polymorphisms on Tac $C_0/D$ in the early stage after heart transplantation. *$p < 0.05$ vs. the CYP3A7 rs2257401 CC group. (**G**) Effect of IL3 rs181781 gene polymorphisms on Tac $C_0/D$ in the early stage after heart transplantation. *$p < 0.05$ vs. the IL3 rs181781 AA group. Data were presented as mean ± SEM. Tac: tacrolimus; $C_0$: blood trough TAC concentration; $C_0/D$: dose-adjusted trough blood concentrations.

Figure 1. The change in mean $C_0/D$ of Tac after heart transplantation

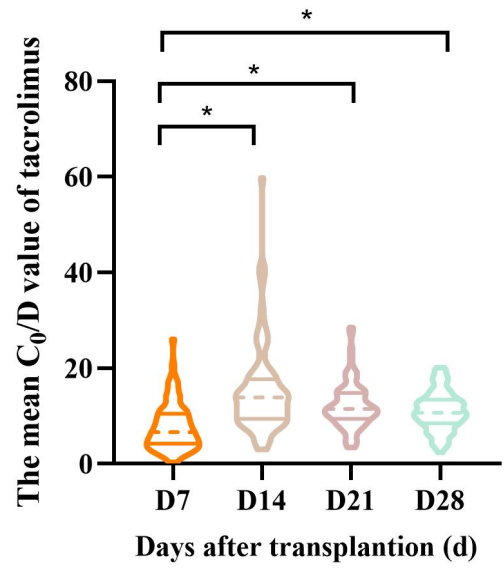

Figure 2. Characterization of genotype distribution of CYP3A4*1G single nucleic acid polymorphisms at different Tac blood concentration distributions.

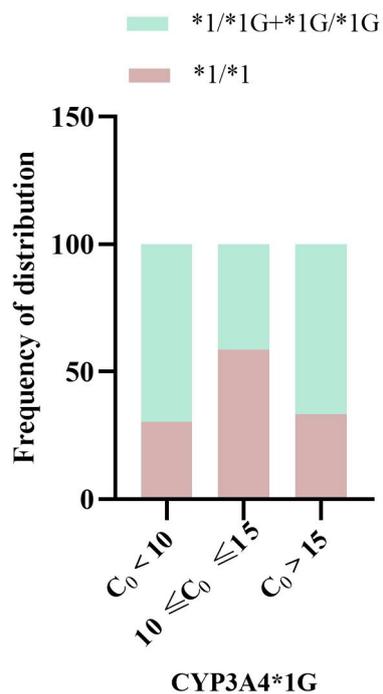

Figure 3. Influence of different genotypes on the Tac $C_0$/D ratio.

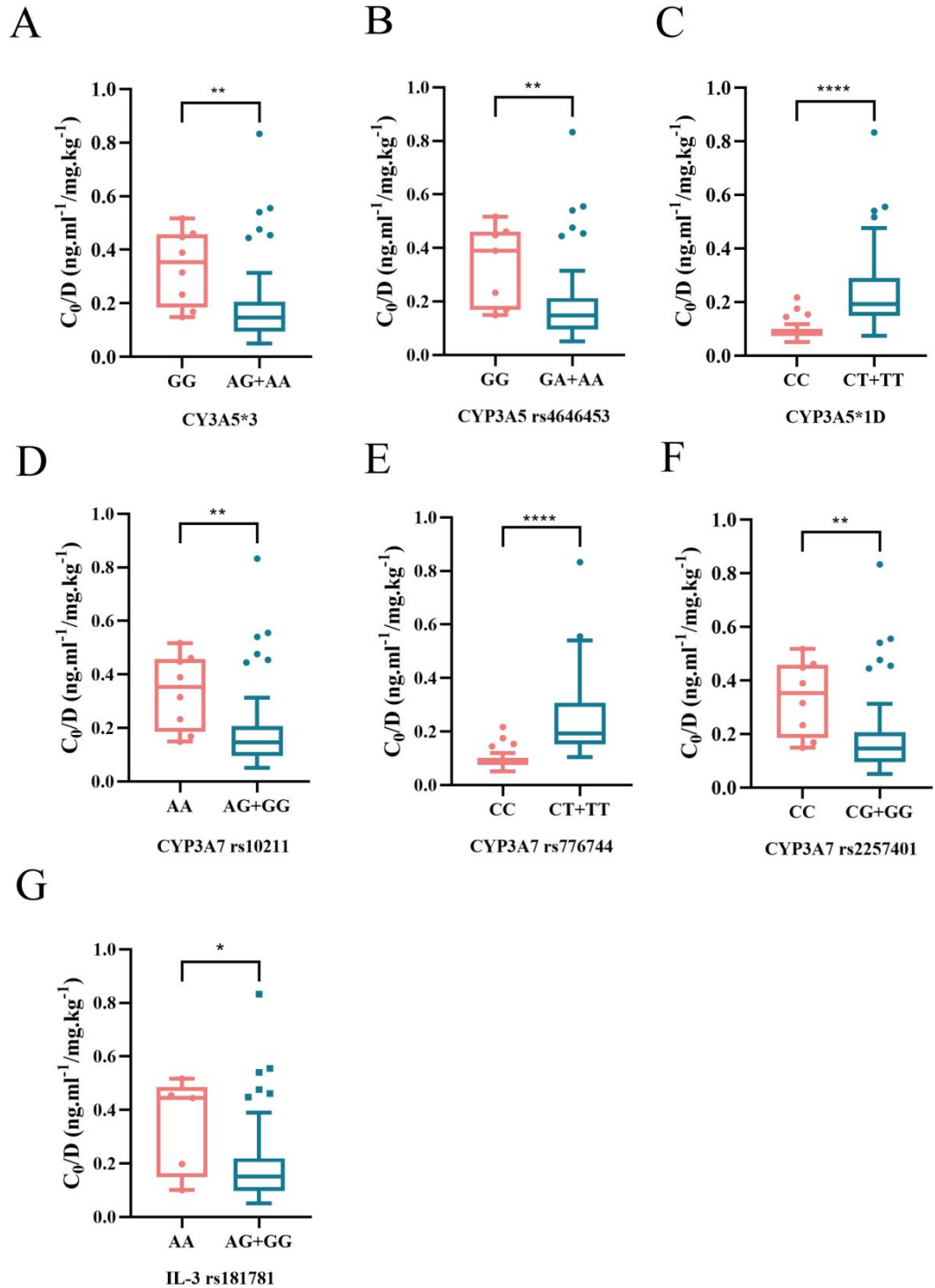